\documentclass[sigconf, nonacm, screen]{acmart}
\AtBeginDocument{%
  }

\setcopyright{cc}
\copyrightyear{2018}
\acmYear{2018}
\acmDOI{XXXXXXX.XXXXXXX}
\acmConference[Conference acronym 'XX]{Make sure to enter the correct
  conference title from your rights confirmation email}{June 03--05,
  2018}{Woodstock, NY}
\acmISBN{978-1-4503-XXXX-X/2018/06}

\begin{document}

\title{Toward a Unified Framework for Collaborative Design of Human-AI Interaction}

\author{Ankur Bhatt}
\orcid{0009-0001-7115-3181}
\affiliation{%
  \institution{TU Dortmund University}
  \city{Dortmund}
  \country{Germany}
}
\affiliation{%
  \institution{Research Center Trustworthy Data Science and Security}
  \city{Dortmund}
  \country{Germany}
}
\email{ankur.bhatt@tu-dortmund.de}

\author{Sven Mayer}
\orcid{0000-0001-5462-8782}
\affiliation{%
  \institution{TU Dortmund University}
  \city{Dortmund}
  \country{Germany}
}
\affiliation{%
  \institution{Research Center Trustworthy Data Science and Security}
  \city{Dortmund}
  \country{Germany}
}
\email{sven.mayer@tu-dortmund.de}

\renewcommand{\shortauthors}{Bhatt et al.}

\begin{abstract}
Human–computer interaction is shifting from screen-based systems to multimodal interfaces where artificial intelligence-powered systems increasingly interpret user intent through speech, gesture, and gaze. Yet users rarely understand how these interpretations are made, compromising trust and control. Existing approaches treat multimodal alignment, explainability, and human agency as separate concerns, leaving critical gaps in transparency and user oversight.
We propose a Human-Artificial Intelligence collaboration framework integrating these three principles as interdependent design requirements: 1) multimodal alignment for accurate intent interpretation, 2) interaction-centric explainability delivering real-time visual, textual, and audio feedback, and 3) agency-preserving mechanisms enabling users to accept, reject, or modify artificial intelligence suggestions at any time.
We presented the framework through two scenarios, collaborative design and extended-reality warehouse robot collaboration, chosen to span differences in time pressure and error reversibility, with the latter situated in a domain where misinterpretation carries documented safety consequences~\cite{yang2025exploring,hostettler2025real}.
This approach reframes collaboration as a continuous interaction property, benefiting designers, researchers, and end users by ensuring that as artificial intelligence systems grow more proactive, user understanding and control remain first-class design properties.
\end{abstract}




\maketitle

\begin{figure}[t]
  \includegraphics[width=\linewidth]{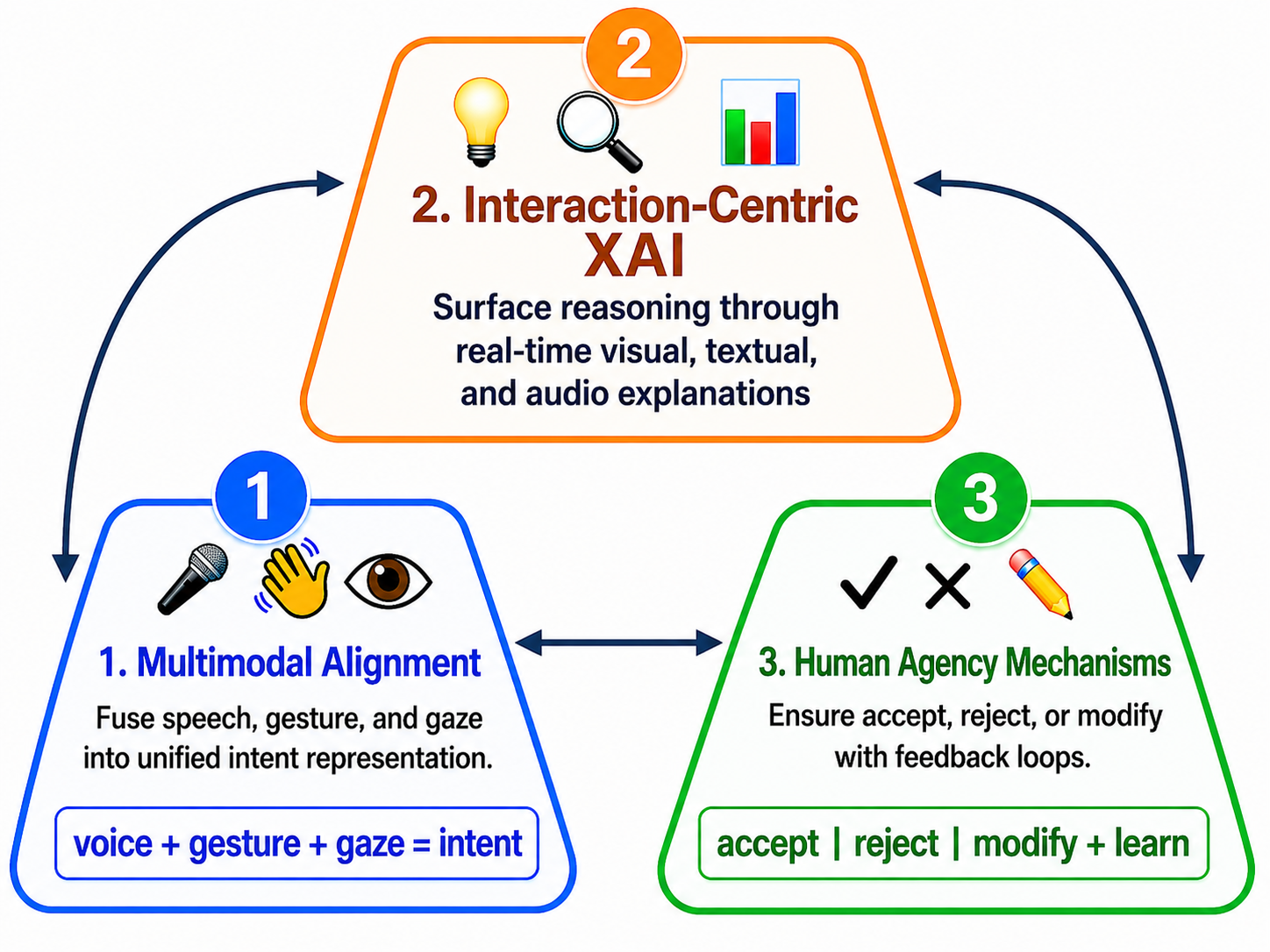}
  \caption{Framework overview showing three interdependent principles for Human-AI collaboration in multimodal systems.}
  \label{fig:framework}
\end{figure}

\section{Introduction}
The integration of artificial intelligence (AI) into interactive systems is changing the way humans engage with technology. Today, AI-powered interfaces span from smart environments and extended reality (XR) systems to online content moderation and human-robot collaboration. These systems increasingly rely on multimodal interaction combining speech, gesture, gaze, and contextual signals to interpret user intent and provide context-aware responses~\cite{cho2025persistent,han2025dynamic,leo2025sensible}. Recent research highlights that AI's involvement in such systems is not limited to prediction accuracy, but includes real-time interpretation and action generation with humans~\cite{cho2025persistent,han2025dynamic,leo2025sensible}. Yet a critical problem persists: users often cannot understand how multimodal AI systems interpret their input, why specific decisions are made, or when they should intervene to correct system behavior. This opacity is especially acute when systems generate proactive actions without making their reasoning explicit~\cite{harari2026proactiveaiadoptionthreatening}, leading to breakdowns in trust and user agency. In warehouse robotics, misinterpreted spatial commands create safety hazards. In content moderation, opaque removal decisions silence legitimate speech, while moderators cannot identify systematic errors without understanding which signals drove classifications. In collaborative design, unexplained changes disrupt creative flow and erode trust. Despite these critical issues, existing approaches treat multimodal alignment, explainability, and user control as separate technical problems, leaving critical gaps precisely where these concerns intersect, and real interaction breakdowns occur. 

Recent work investigates multimodal intent grounding advances fusion techniques for speech, gesture, and gaze~\cite{cho2025persistent, han2025dynamic, leo2025sensible} to address the prevalent issues. At the same time, explainable AI (XAI) develops methods for surfacing model reasoning to support trust and accountability in algorithmic decisions~\cite{lundberg2017unifiedapproachinterpretingmodel, ribeiro2016whyshould, samek2019explainable}. 

We propose a Human-AI collaboration framework integrating multimodal alignment, interaction-centric explainability, and agency-preserving control as interdependent components. This framework addresses the problems of multimodal alignment, explainability, and human agency together: the AI remains transparent about the signals driving its decisions and enables users to accept, reject, or modify suggestions at any point, unlike approaches adding XAI post-hoc~\cite{lundberg2017unifiedapproachinterpretingmodel,ribeiro2016whyshould,samek2019explainable} or emphasizing interpretation accuracy over user understanding~\cite{cho2025persistent,han2025dynamic,leo2025sensible}. We treat these three concerns as a single design problem. Our key insight is mutual reinforcement: alignment quality determines what XAI can surface, explanation quality determines how meaningfully users exercise agency, and agency mechanisms feeding corrections back improve future alignment. Our framework generates explanations alongside decisions and enables users to accept, reject, or modify suggestions at any point. The system surfaces which specific signals (e.g., "voice='resize' + gesture $85\%$ match + gaze confirmation") drove each decision through visual, textual, and audio feedback, treating corrections as preference data shaping future behavior.

Through scenario-based analysis across collaborative user-interface design and 
XR robot collaboration, we demonstrate how this integrated approach 
addresses interaction breakdowns that occur when principles are treated in isolation. Our framework reveals three  findings:
\begin{enumerate}
    \item Alignment quality determines downstream transparency: when fusion correctly identifies intent by combining speech + gesture + gaze, users receive precise explanations (e.g., 'voice=resize + gesture 85\% match'), but failed alignment produces misleading rationales.
    \item Real-time explanations enable users to correct AI interpretations at the moment of decision, transforming interaction from reactive debugging to proactive negotiation.
    \item Agency mechanisms treating corrections as preference signals create co-adaptive learning loops that refine future behavior while preserving human authority.
\end{enumerate}
Our scenarios illustrate how integrated explanation-agency mechanisms address misinterpretation-related errors by making ambiguity visible at decision points,  enabling users to correct or refine AI behavior before it propagates. This work demonstrates that as AI systems become more proactive, understanding and control 
must be first-class design properties rather than features added afterward.

\section{Related Work}
Our framework lies at the intersection of multimodal interaction, human-centered and XAI, and emerging frameworks for interactive AI agents. While each of these areas has been studied extensively, prior work has largely addressed them separately.
\subsection{Multimodal Interaction and Interpretation}
Research in multimodal interaction has long demonstrated that combining modalities such as speech, gesture, and gaze can improve robustness, efficiency, and user satisfaction. \citet{oviatt2007multimodal} foundational work shows that multimodal interfaces enable users to distribute cognitive load across complementary channels, reducing errors compared to unimodal input alone ~\cite{oviatt2007multimodal}. Subsequent systems have explored practical mechanisms for fusing speech and gesture, often treating modalities as parallel input streams that are integrated at later stages of processing. More recent work has advanced multimodal intent grounding in interactive systems. \citet{cho2025persistent} introduces Persistent Assistant, which maintains continuity across interactions by grounding user intent through multimodal cues and feedback~\cite{cho2025persistent}. Similarly, \citet{han2025dynamic} proposed a Dynamic Bayesian Network framework to model multimodal context and uncertainty in interactive settings~\cite{han2025dynamic}. In immersive and spatial interfaces, \citet{leo2025sensible} demonstrated how proactive AR agents can interpret multimodal signals while remaining unobtrusive~\cite{leo2025sensible}. While these systems show the benefits of multimodal fusion, they primarily emphasize accurate interpretation and system responsiveness, with less attention to how users understand or interrogate the AI's internal reasoning.

\begin{figure*}
  \includegraphics[width=\linewidth]{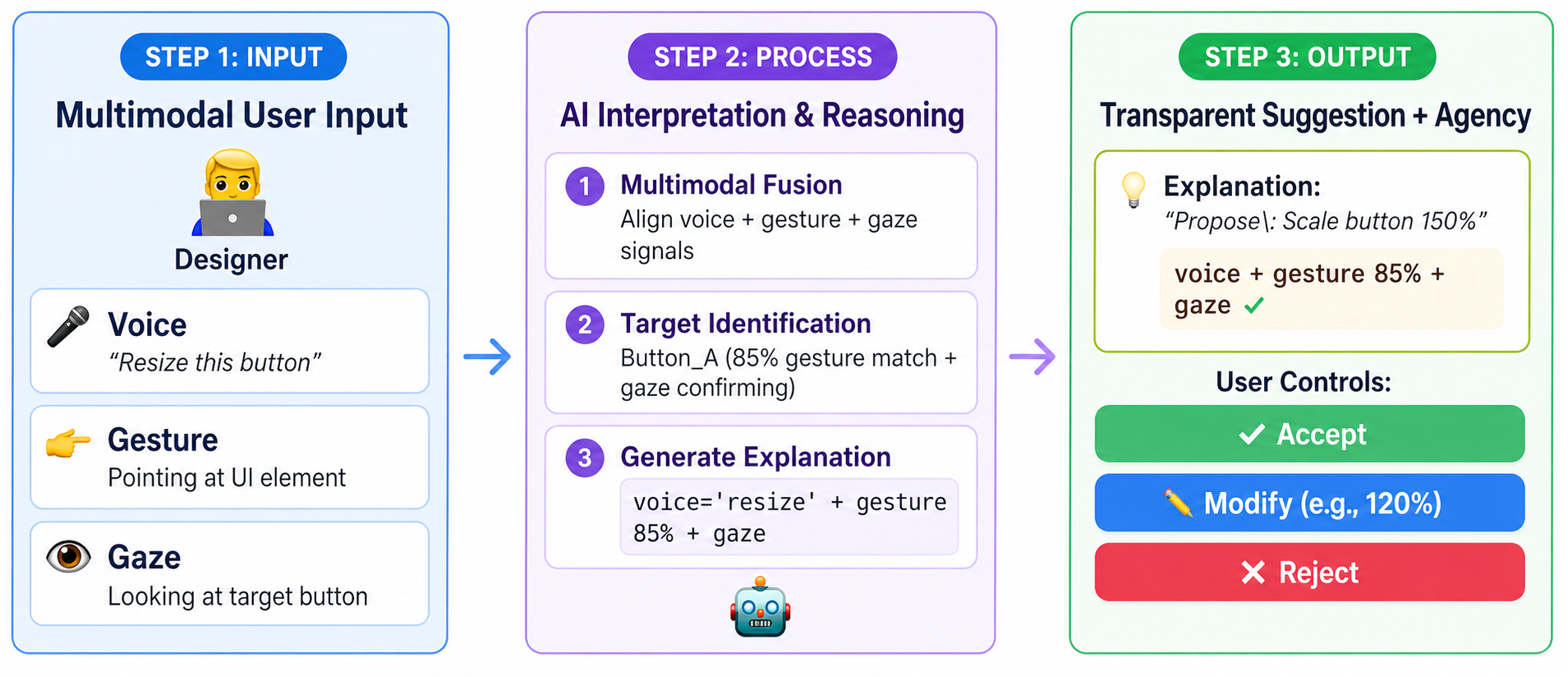}
  \caption{Illustrative scenario demonstrating the framework via Human–AI collaboration.}
  \label{fig:scenario}
\end{figure*}

\subsection{Explainable and Interpretable AI in Interaction}
XAI has produced a wide range of techniques for making model behavior more transparent. Model-agnostic approaches such as LIME~\cite{ribeiro2016whyshould} and SHAP~\cite{lundberg2017unifiedapproachinterpretingmodel} enable post hoc explanations of predictions from complex models, while broader surveys and frameworks emphasize interpretability as a prerequisite for trust and accountability~\cite{samek2019explainable}. However, much of this work targets static prediction settings or single-modal inputs, such as images or tabular data. Within HCI, XAI has increasingly been framed as an interaction design problem rather than a purely technical one. \citet{amreshi2019guidelines} outline guidelines for Human–AI Interaction that emphasize transparency, intelligibility, and appropriate user control~\cite{amreshi2019guidelines}. In multimodal and immersive contexts, \citet{macgowan2025design} presented the Transparent, Interpretable, and Multimodal (TIM) Augmented Reality (AR) assistant, which integrates interpretable visual and multimodal feedback into an interactive AR system~\cite{macgowan2025design}. Recent proposals, such as ACE, argue that explanations should be action-oriented and embedded into user workflows, particularly for multimodal AI assistants powered by large language models~\cite{watkins2025aceactioncontrolexplanations}. Despite these advances, XAI mechanisms are often added after intent interpretation, rather than being tightly coupled to multimodal alignment itself. Only a small body of work has begun to explicitly connect multimodal interaction with XAI. \citet{viswanath2025enhancing} demonstrates how robots can leverage multimodal context representations to produce more explainable behavior. Recent machine learning approaches, such as interpretable multimodal mixture-of-experts models, further aim to expose how different modalities contribute to system decisions~\cite{xin2025i2moeinterpretablemultimodalinteractionaware}. Vision papers on multisensory intelligence argue for tighter integration between sensing, reasoning, and explanation across modalities~\cite{liang2026visionmultisensoryintelligencesensing}.

\subsection{Human Agency and Human-in-the-Loop AI}
Human-centered AI research stresses the importance of preserving user agency, especially as AI systems become more autonomous. Design principles for keeping humans "in the loop" highlight the need for oversight, reversibility, and meaningful intervention~\cite{amreshi2019guidelines}. These concerns are addressed in broader policy frameworks such as the "\href{https://digital-strategy.ec.europa.eu/en/library/ethics-guidelines-trustworthy-ai}{Trustworthy AI}" guidelines, which emphasize transparency and user control in AI-assisted systems. In interactive AI agents, recent systems increasingly support incremental or mixed-initiative interaction rather than full automation. For example, proactive agents in AR and multimodal assistants are designed to suggest actions while allowing users to accept, reject, or modify them~\cite{cho2025persistent,leo2025sensible}. However, these systems may rely on implicit signals of user satisfaction or disengagement, leaving limited opportunities for explicit correction or negotiation. As a result, mismatches between user intent and system behavior can still erode trust, particularly when multimodal cues are ambiguous or conflicting.

\section{Methodology: Three Guiding Principles}
This work adopts a conceptual and design-oriented approach to explore how humans and AI systems can collaborate more effectively in multimodal interactive contexts. Rather than starting from a fixed implementation or empirical study, we focus on designing a framework and examining its implications through concrete scenarios. We present this as a conceptual framework deliberately; our goal is to establish the design space and identify the interdependencies between principles before empirical validation, which we position as future work. The three principles are not independent design goals but mutually reinforcing components: the quality of multimodal alignment directly determines what the XAI layer can surface, and the usefulness of explanations directly determines how meaningfully users can exercise agency, as shown in \autoref{fig:framework}. A failure in any one principle propagates through the others, which is precisely why they must be treated as a unified framework rather than separate concerns.

\subsection{Multimodal Alignment}
The first principle addresses how multimodal input is captured and aligned in practice. The system incorporates speech, gesture, and eye gaze as complementary signals for interpreting user intent. Speech input is processed using natural language understanding (NLU) models to extract commands and contextual cues, while gestures are tracked via motion sensors, depth cameras, or hand-tracking devices to identify potential referents and actions. Eye gaze contributes information about user attention and focus, though we treat gaze as a probabilistic rather than definitive indicator of intent. These heterogeneous signals are aligned using multimodal fusion techniques such as shared semantic representations or transformer-based embeddings, so that meaning emerges from their combination rather than from any single modality in isolation. Critically, the quality of this alignment is the foundation upon which the entire framework rests; if intent is misinterpreted at this stage, no amount of explanation or agency mechanism can fully recover user trust. This makes alignment not merely a technical preprocessing step but a first-class design concern that must be made visible to downstream components.

\subsection{Interaction-Centric XAI}
The second principle addresses how the system communicates its reasoning to users in real time. The XAI layer externalizes the AI's reasoning through interaction, using visual, textual, and auditory explanations tailored to the context. Visual cues may highlight interface elements or parameters that influenced a recommendation, while short textual explanations summarize the underlying rationale in accessible terms. Audio feedback complements these explanations in immersive or hands-free contexts such as XR environments. Importantly, explanations are treated as part of the interaction design rather than as post-hoc justifications, as they are generated alongside the decision, not after it. This principle depends directly on the quality of multimodal alignment: a well-aligned interpretation produces an explanation grounded in specific, identifiable signals, while a poorly aligned interpretation produces an explanation that is either misleading or too vague to be actionable. Conversely, the XAI layer shapes what agency mechanisms can offer a user who understands why a suggestion was made is far better positioned to accept, reject, or meaningfully modify it than one who receives only a confidence score.

\subsection{Human Agency Mechanisms}
The third principle addresses how users retain meaningful control over AI behavior throughout the interaction. Users retain the ability to accept, reject, or modify AI suggestions at any point, ensuring that decision-making authority remains with the human. User actions are fed back into the system as signals for learning and adaptation, creating a co-adaptive loop in which the AI refines its behavior while remaining responsive to explicit user control. This principle is the most vulnerable to failure when the preceding two principles are weak. If alignment is poor, users may be correcting the wrong interpretation entirely, and if explanations are inadequate, users cannot make informed correction decisions. This approach deliberately prioritizes interpretability and user oversight over maximal automation, reflecting concerns around over-reliance and loss of agency in interactive AI systems. The co-adaptive loop also raises an important design question that we surface for future empirical investigation: how should the system balance learned user preferences against new contextual signals, particularly when the two conflict? To illustrate how these components interact, we developed two scenario-based demonstrations in Section~\ref{sub:sce}.


\section{Illustrative Scenarios}
\label{sub:sce}
We illustrate the framework through two scenarios chosen to span differences in time pressure and error reversibility: a collaborative user-interface design task and an XR warehouse robot collaboration task.
\paragraph{Without the framework} The AI would either resize the wrong element due to an ambiguous voice command alone, or remain silent without justification, leaving the designer confused about what changed and why. With the framework, multimodal alignment fuses the voice command, pointing gesture, and gaze confirmation to correctly identify the target button. The XAI layer surfaces the reasoning,  "Proposed: Scale button $150\%$ (voice='resize' + gesture $85\%$ match + gaze confirmation)", so the designer understands exactly why that button was selected as shown in \autoref{fig:scenario}. Agency mechanisms then allow the designer to accept, modify the scale value, or reject entirely. A challenge remains: when gesture and gaze conflict, for instance, the user points at one element while looking at another, the system must decide which signal to prioritize, and that decision should itself be made visible to the user. 

\paragraph{Without the framework} A robot acting on speech alone in a noisy warehouse environment may misinterpret "stop there" without spatial grounding, stopping at an incorrect location with no explanation offered to the operator. This is particularly dangerous in time-sensitive logistics environments where incorrect positioning can cause safety incidents. With the framework, the robot fuses the speech command with gaze data indicating the $2m$ zone, producing a spatially grounded stop decision. The AR overlay makes the reasoning transparent in real time: "Speech=halt + gaze=$2m$ zone; safe stop at $1.5m$". When the operator overrides with "$1m$ instead," the system replans and confirms via audio, "Scoped to $1m$ per override," closing the feedback loop explicitly. The agency mechanism here is critical: the operator is not just correcting the robot but teaching it, as the override is logged as a preference signal. Residual challenges include latency; in fast-moving warehouse environments, the explanation overlay must appear quickly enough to be actionable, and the question of how much explanation detail is appropriate under time pressure without becoming a distraction.

\section{Limitations} 
This work presents a conceptual framework without empirical validation; we have not implemented a working system or conducted user studies to confirm that (a) the three principles mutually reinforce each other in practice, (b) real-time explanations improve user trust, or (c) the co-adaptive loop refines behavior without amplifying biases or creating filters when users provide contradictory corrections. Our framework prioritizes transparency over efficiency, making it inappropriate for high-frequency, less critical interactions where users prefer speed over explanation, and we have not validated whether explanations improve or degrade performance in time-pressured environments like Scenario 2's warehouse. The framework assumes reliable multimodal sensing, but gaze tracking degrades in bright sunlight or with glasses, gesture recognition struggles in cluttered spaces, and speech recognition fails under industrial noise. When sensors fail, poor alignment yields misleading explanations, corrupting the co-adaptive loop. Scenario 1 raises gesture-gaze conflicts but provides no resolution mechanism. Implementation challenges include achieving low-latency explanation generation for interactive responsiveness, managing privacy concerns from continuous tracking, and tuning explanation verbosity per domain and user. The framework may not suit safety-critical systems requiring split-second decisions or privacy-sensitive contexts rejecting continuous observation.

\section{Summary}
This position paper proposes that effective interaction with AI-powered multimodal systems requires treating multimodal interpretation, XAI, and human agency as inseparable design concerns. As AI agents increasingly participate in interactions such as interpreting speech, gesture, and gaze and acting on behalf of users, failures in transparency or control can undermine understanding and trust. Addressing these challenges requires moving beyond isolated technical solutions toward interaction designs that support negotiation, feedback, and correction as interaction unfolds. We propose a Human–AI collaboration framework that addresses these concerns and uses scenario-based analysis to illustrate where interaction breakdowns occur and how they can be mitigated. Rather than eliminating ambiguity, the framework emphasizes making ambiguity visible and manageable through explanation and user intervention. We argue that this perspective is essential as AI systems become more proactive and embedded in everyday interfaces. This work highlights the need for human-centered approaches that prioritize understanding and agency alongside capability, and it offers a foundation for future empirical and design-oriented research on collaborative interactive AI systems.

\begin{acks}
This work has been partly supported by the Research Center Trustworthy Data Science and Security (\href{https://rc-trust.ai}{https://rc-trust.ai}), one of the Research Alliance centers within the UA Ruhr (\href{https://uaruhr.de}{https://uaruhr.de}).
\end{acks}

\bibliographystyle{ACM-Reference-Format}
\bibliography{bibliography}

@misc{harari2026proactiveaiadoptionthreatening,
      title={Proactive AI Adoption can be Threatening: When Help Backfires}, 
      author={Dana Harari and Ofra Amir},
      year={2026},
      eprint={2509.09309},
      archivePrefix={arXiv},
      primaryClass={cs.HC},
      url={https://arxiv.org/abs/2509.09309}, 
}

@inproceedings{yang2025exploring,
author = {Yang, Xiliu and Pathmanathan, Nelusa and Zabel, Sarah and Amtsberg, Felix and Otto, Siegmar and Kurzhals, Kuno and Sedlmair, Michael and Menges, Achim},
title = {Exploring the Use of Augmented Reality for Multi-human-robot Collaboration with Industry Users in Timber Construction},
year = {2025},
isbn = {9798400713958},
publisher = {Association for Computing Machinery},
address = {New York, NY, USA},
url = {https://doi.org/10.1145/3706599.3720104},
doi = {10.1145/3706599.3720104},
booktitle = {Proceedings of the Extended Abstracts of the CHI Conference on Human Factors in Computing Systems},
articleno = {272},
numpages = {8},
location = {
},
series = {CHI EA '25}
}

@inproceedings{hostettler2025real,
author = {Hostettler, Damian and Mayer, Simon and Albert, Jan Liam and Jenss, Kay Erik and Hildebrand, Christian},
title = {Real-Time Adaptive Industrial Robots: Improving Safety And Comfort In Human-Robot Collaboration},
year = {2025},
isbn = {9798400713941},
publisher = {Association for Computing Machinery},
address = {New York, NY, USA},
url = {https://doi.org/10.1145/3706598.3713889},
doi = {10.1145/3706598.3713889},
booktitle = {Proceedings of the 2025 CHI Conference on Human Factors in Computing Systems},
articleno = {908},
numpages = {16},
location = {
},
series = {CHI '25}
}

@misc{liang2026visionmultisensoryintelligencesensing,
      title={A Vision for Multisensory Intelligence: Sensing, Science, and Synergy}, 
      author={Paul Pu Liang},
      year={2026},
      eprint={2601.04563},
      archivePrefix={arXiv},
      primaryClass={cs.LG},
      url={https://arxiv.org/abs/2601.04563}, 
}

@inproceedings{leo2025sensible,
author = {Lee, Geonsun and Xia, Min and Numan, Nels and Qian, Xun and Li, David and Chen, Yanhe and Kulshrestha, Achin and Chatterjee, Ishan and Zhang, Yinda and Manocha, Dinesh and Kim, David and Du, Ruofei},
title = {Sensible Agent: A Framework for Unobtrusive Interaction with Proactive AR Agents},
year = {2025},
isbn = {9798400720376},
publisher = {Association for Computing Machinery},
address = {New York, NY, USA},
url = {https://doi.org/10.1145/3746059.3747748},
doi = {10.1145/3746059.3747748},
booktitle = {Proceedings of the 38th Annual ACM Symposium on User Interface Software and Technology},
articleno = {119},
numpages = {22},
location = {
},
series = {UIST '25}
}

@misc{watkins2025aceactioncontrolexplanations,
      title={ACE, Action and Control via Explanations: A Proposal for LLMs to Provide Human-Centered Explainability for Multimodal AI Assistants}, 
      author={Elizabeth Anne Watkins and Emanuel Moss and Ramesh Manuvinakurike and Meng Shi and Richard Beckwith and Giuseppe Raffa},
      year={2025},
      eprint={2503.16466},
      archivePrefix={arXiv},
      primaryClass={cs.HC},
      url={https://arxiv.org/abs/2503.16466}, 
}

@inproceedings{cho2025persistent,
author = {Cho, Hyunsung and Fashimpaur, Jacqui and Sendhilnathan, Naveen and Browder, Jonathan and Lindlbauer, David and Jonker, Tanya R. and Todi, Kashyap},
title = {Persistent Assistant: Seamless Everyday AI Interactions via Intent Grounding and Multimodal Feedback},
year = {2025},
isbn = {9798400713941},
publisher = {Association for Computing Machinery},
address = {New York, NY, USA},
url = {https://doi.org/10.1145/3706598.3714317},
doi = {10.1145/3706598.3714317},
booktitle = {Proceedings of the 2025 CHI Conference on Human Factors in Computing Systems},
articleno = {59},
numpages = {19},
location = {
},
series = {CHI '25}
}

@misc{xin2025i2moeinterpretablemultimodalinteractionaware,
      title={I2MoE: Interpretable Multimodal Interaction-aware Mixture-of-Experts}, 
      author={Jiayi Xin and Sukwon Yun and Jie Peng and Inyoung Choi and Jenna L. Ballard and Tianlong Chen and Qi Long},
      year={2025},
      eprint={2505.19190},
      archivePrefix={arXiv},
      primaryClass={cs.LG},
      url={https://arxiv.org/abs/2505.19190}, 
}

@inproceedings{han2025dynamic,
author = {Han, Violet Yinuo and Wang, Tianyi and Cho, Hyunsung and Todi, Kashyap and Fernandes, Ajoy Savio and Levi, Andre and Zhang, Zheng and Grossman, Tovi and Ion, Alexandra and Jonker, Tanya R.},
title = {A Dynamic Bayesian Network Based Framework for Multimodal Context-Aware Interactions},
year = {2025},
isbn = {9798400713064},
publisher = {Association for Computing Machinery},
address = {New York, NY, USA},
url = {https://doi.org/10.1145/3708359.3712070},
doi = {10.1145/3708359.3712070},
booktitle = {Proceedings of the 30th International Conference on Intelligent User Interfaces},
pages = {54–69},
numpages = {16},
location = {
},
series = {IUI '25}
}

@article{oviatt2007multimodal,
  title={Multimodal interfaces},
  author={Oviatt, Sharon},
  journal={The human-computer interaction handbook},
  pages={439--458},
  year={2007},
  publisher={CRC press}
}

@inproceedings{ribeiro2016whyshould,
author = {Ribeiro, Marco Tulio and Singh, Sameer and Guestrin, Carlos},
title = {"Why Should I Trust You?": Explaining the Predictions of Any Classifier},
year = {2016},
isbn = {9781450342322},
publisher = {Association for Computing Machinery},
address = {New York, NY, USA},
url = {https://doi.org/10.1145/2939672.2939778},
doi = {10.1145/2939672.2939778},
booktitle = {Proceedings of the 22nd ACM SIGKDD International Conference on Knowledge Discovery and Data Mining},
pages = {1135–1144},
numpages = {10},
location = {San Francisco, California, USA},
series = {KDD '16}
}

@misc{lundberg2017unifiedapproachinterpretingmodel,
      title={A Unified Approach to Interpreting Model Predictions}, 
      author={Scott Lundberg and Su-In Lee},
      year={2017},
      eprint={1705.07874},
      archivePrefix={arXiv},
      primaryClass={cs.AI},
      url={https://arxiv.org/abs/1705.07874}, 
}

@book{samek2019explainable,
author = {Samek, Wojciech and Montavon, Gregoire and Vedaldi, Andrea and Hansen, Lars Kai and Muller, Klaus-Robert},
title = {Explainable AI: Interpreting, Explaining and Visualizing Deep Learning},
year = {2019},
isbn = {3030289532},
publisher = {Springer Publishing Company, Incorporated},
edition = {1st}
}

@inproceedings{amreshi2019guidelines,
author = {Amershi, Saleema and Weld, Dan and Vorvoreanu, Mihaela and Fourney, Adam and Nushi, Besmira and Collisson, Penny and Suh, Jina and Iqbal, Shamsi and Bennett, Paul N. and Inkpen, Kori and Teevan, Jaime and Kikin-Gil, Ruth and Horvitz, Eric},
title = {Guidelines for Human-AI Interaction},
year = {2019},
isbn = {9781450359702},
publisher = {Association for Computing Machinery},
address = {New York, NY, USA},
url = {https://doi.org/10.1145/3290605.3300233},
doi = {10.1145/3290605.3300233},
booktitle = {Proceedings of the 2019 CHI Conference on Human Factors in Computing Systems},
pages = {1–13},
numpages = {13},
location = {Glasgow, Scotland Uk},
series = {CHI '19}
}

@article{viswanath2025enhancing,
  doi = {10.5281/ZENODO.14930029},
  
  url = {https://zenodo.org/doi/10.5281/zenodo.14930029},
  
  author = {Viswanath, Anargh and Veeramacheneni, Lokesh and Buschmeier, Hendrik},
  
  title = {Enhancing Explainability with Multimodal Context Representations for Smarter Robots},
  
  publisher = {Zenodo},
  
  year = {2025},
  
  copyright = {Creative Commons Attribution 4.0 International}
}

@ARTICLE{macgowan2025design,
  author={McGowan, Erin and Rulff, Joao and Castelo, Sonia and Wu, Guande and Chen, Shaoyu and Lopez, Roque and Steers, Bea and Roman, Iran R. and Dias, Fábio F. and Qian, Jing and Solunke, Parikshit and Middleton, Michael and McKendrick, Ryan and Silva, Cláudio T.},
  journal={IEEE Computer Graphics and Applications}, 
  title={Design and Implementation of the Transparent, Interpretable, and Multimodal (TIM) AR Personal Assistant}, 
  year={2025},
  volume={45},
  number={1},
  pages={28-42},
  doi={10.1109/MCG.2025.3549696}}

\end{document}